\documentclass[aps,pra,reprint,groupedaddress]{revtex4-2}

\usepackage{amsmath,amssymb}
\usepackage{graphicx}
\usepackage{bm} 
\usepackage{hyperref}

\renewcommand{\d}{{\rm d}}
\newcommand{\e}{{\rm e}}
\newcommand{\imai}{{\rm i}}

\begin{document}
\title{Nonresonant two-level transitions: Insights from quantum thermodynamics}
\author{Andreas Wacker}
\email{Andreas.Wacker@teorfys.lu.se}
\affiliation{Mathematical Physics and NanoLund, Lund University, Box 118, 22100 Lund, Sweden}
\date{11 January 2022, accepted by Physical Review A}
\begin{abstract}
Based on concepts from quantum thermodynamics the two-level system coupled to a single electromagnetic mode is analyzed. Focusing on the case of detuning, where the mode frequency does not match the transition frequency, effective energies are derived for the levels and the photon energy. It is shown that these should be used for energy exchange with fermionic and bosonic reservoirs in the steady state in order to achieve a thermodynamically consistent description. While recovering known features such as frequency pulling or Bloch gain, this sheds light on their thermodynamic background and allows for a coherent understanding.
\end{abstract}
\maketitle
\section{Introduction}
Two-level systems are the paradigm for the interaction of matter with
light. Typically, one considers light frequencies $\omega/2\pi$, where
the photon energy $\hbar\omega$ matches the energy difference
$E_u-E_l$ between the upper (index $u$) and lower (index $l$)
level. However, it is well known, that optical transitions are
broadened due to the finite lifetimes of levels and photons, so that a
certain width of frequencies can be emitted or absorbed. A
straightforward question is, how energy balance is satisfied for a
finite detuning $\hbar\Delta=\hbar\omega +E_l-E_u$.

Detuning is known to have a variety of practical consequences. E.g., it results in frequency pulling \cite{SiegmanBook1986} for lasers and Bloch gain for inter-subband transitions in semiconductor heterostructures \cite{WillenbergPRB2003,TerazziNatPhys2007}. From a more fundamental point of view there had been discussions on the 
thermodynamic consistency \cite{BoukobzaPRL2007} for the archetypal Scovil\&Schulz-DuBois maser \cite{ScovilPRL1959}.
Detuned transitions play also an important role for certain gate operations
on Qubits in quantum information \cite{RoosPRA2004,SaffmanRMP2010}.

Here, the issue of detuning is studied from a quantum-thermodynamic
\cite{BinderBook2018,StreltsovRevModPhys2017,BenentiPhysRep2017,DeffnerBook2019}
perspective. Assuming local couplings \cite{LevyEPL2014,HoferNJP2017,GonzalezOpenSystInfoDyn2017} of the two levels with separate reservoirs, the energy balance in the steady state allows to identify
effective energies for the levels and the
electromagnetic mode. These differ from the bare energies by a
fraction of the detuning which is proportional to the contribution to
the total broadening, see
Eqs.~(\ref{EqEffectiveEnergies},\ref{EqEffectiveEnergiesQuant}) for
the classical and quantum treatment, respectively. Applying the effective energies for the reservoir  transitions it is shown that both the first and second law
of thermodynamics are satisfied.

\begin{figure}
	\includegraphics[width=7cm]{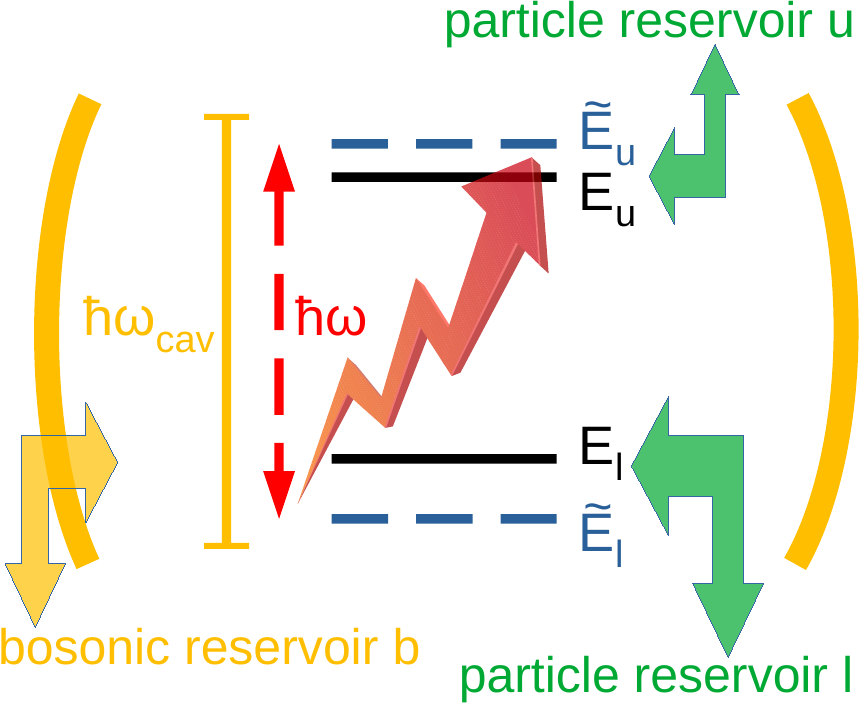}
	\caption{\label{fig:System} Sketch of the two-level system
          (horizontal black full lines) coupled to particle reservoirs
          and a single mode electromagnetic field. If the photon
          energy $\hbar\omega$ does not match the energy difference
          between the levels, the analysis of energy fluxes provides
          effective energies (dashed blue horizontal lines), see
          Eq.~\eqref{EqEffectiveEnergies}, which satisfy energy
          conservation $\tilde{E}_u-\tilde{E}_l=\hbar\omega$. For a
          quantum treatment of the cavity mode coupled to a bosonic
          reservoir, the light comes in portions with an effective
          energy $\tilde{E}_\textrm{ph}=\hbar\omega$, see
          Eq.~\eqref{EqEffectiveEnergiesQuant} which differs from the
          photon energy of the empty cavity $\hbar\omega_\textrm{cav}$
          by a fraction of the total detuning
          $\hbar\Delta_\textrm{cav}=\hbar\omega_\textrm{cav}+\tilde{E}_l-\tilde{E}_u$. }
\end{figure}

In this work, fermionic systems are considered, where the upper/lower
level is connected to a specific reservoir with electrochemical
potential $\mu_{u/l}$, as depicted in Fig.~\ref{fig:System}. Such a
set-up was recently realized experimentally with high conversion
efficiency for microwaves \cite{KhanNatCommun2021}. It is a prototype
model system for many important devices such as light emitting diodes
(LEDs), semiconductor lasers, or semiconductor solar cells, where the
upper/lower level corresponds to conduction/valence band states
respectively. Related results for the states of an atom coupled to
bosonic reservoirs have been presented in \cite{KalaeePRA2021}
restricting to a classical field, which resolved the issues addressed in \cite{BoukobzaPRL2007}.

This article is organized as follows: Sec.~\ref{SecThermo} briefly summaries the general concepts from quantum thermodynamics applied. The heart of the article is Sec.~\ref{Sec2Level}, where the two-level system is carefully analysed using detailed calculations presented in appendix \ref{SecDetailClassics} and \ref{SecDetailQuant} for the classical and quantum treatment of the electromagnetic mode, respectively.
  Sec.~\ref{SecTwoLevelSemiclassical} and Sec.~\ref{SecBloch} consider the frequency pulling of lasers and the Bloch gain for intersubband transitions. For these examples it is shown that the effective energies introduced here provide the same features as detailed microscopic calculations performed before. Finally, Appendix \ref{SecGeneralFermion} details how the effective energies can be generalised to arbitrary systems with fermionic baths.

\section{General thermodynamic point of view}\label{SecThermo}
We consider the system (e.g. the two-level system) in connection with reservoirs using the  common quantum-thermodynamics treatment. Let $\dot{U}_\alpha$ be the energy flow  from the reservoir $\alpha$ into the system and $\dot{U}_\textrm{opt}$  the energy flow from the optical field into the system. Energy conservation implies in the steady state
\begin{equation}
\sum_\alpha 
\dot{U}_\alpha+\dot{U}_\textrm{opt}=0\, .
\label{EqEnergySteadyState}
\end{equation}
The entropy production in reservoir $\alpha$ reads
\begin{equation}
\frac{\d S_\alpha}{\d t}= -\frac{\dot{U}_\alpha}{T_\alpha}+\dot{N}_{\alpha}\frac{\mu_\alpha}{T_\alpha}\, ,
\label{EqEntropyLeadAlpha}
\end{equation} 
where $\dot{N}_{\alpha}$ is the particle transfer from the reservoir $\alpha$ into the system. Then the second law of thermodynamics in the form of positive definite entropy production in the steady state requires
\begin{equation}
\sum_\alpha\frac{\d S_\alpha}{\d t}+ \frac{\d S_\textrm{opt}}{\d t}\ge 0\, .
\label{EqEntropySteadyState}
\end{equation}

A standard quantum kinetic treatment is provided by the time evolution of the reduced density operator $\hat{\rho}$ of the system
\begin{equation}
\frac{\d\hat{\rho}}{\d t}= \frac{1}{\imai\hbar} \left[\hat{H}_S,\hat{\rho}\right]+\sum_\alpha \mathcal{L}_\alpha[\hat{\rho}] \, ,
\label{EqMasterEquation}
\end{equation}
where $\hat{H}_S$ the system Hamilton operator and $\mathcal{L}_\alpha$ describes the coupling to reservoir $\alpha$ within a
Markovian treatment, see e.g. \cite{AlickiBook2018}. Following well established standard procedure \cite{PuszCommMathPhys1978,SpohnBook1978,AlickiJPA1979} which is commonly used for the description of nanoscale engines, see e.g. \cite{KosloffEntropy2017}, the energy and particle flows from the reservoirs into the system are 
\begin{equation}
\dot{U}_\alpha= \textrm{Tr}\{\hat{H}_S\mathcal{L}_\alpha[\hat{\rho}]\}
\qquad\textrm{and }
\dot{N}_\alpha= \textrm{Tr}\{\hat{N}_S\mathcal{L}_\alpha[\hat{\rho}]\}\, ,
\label{EqEnergyLeadAlpha}
\end{equation}
respectively, where $\hat{N}_S$ is the number operator of the system.
The corresponding power (work per time) transferred to the system reads 
\begin{equation}
P_S= \textrm{Tr}\left\{\hat{\rho}\frac{\partial\hat{H}_S}{\partial t}\right\}
\label{EqWQquantumFull}
\end{equation}
which requires an absolute time-dependence of the Hamiltonian, as given by a classical optical field. It matches exactly the total rate of work $\bm{j}\cdot \bm{\mathcal E}$ done by a classical electromagnetic field \cite{JacksonBook1998}, where $\bm{j}$ is the current density and $\bm{\mathcal E}$ is the electric field.

Such thermodynamic considerations have been frequently performed for the analysis of optoelectronic systems such as light emitting diodes (LEDs), Lasers, and solar cells, see e.g.,  \cite{DorfmanAIPConfProc2011,AlickiAnnPhys2017} with a highlight on different aspects. The focus of this work is to use Eqs.~(\ref{EqEnergySteadyState},\ref{EqEntropySteadyState}) as necessary conditions for the choice of  $\mathcal{L}_\alpha$ in the construction of master equations. As shown below, this has distinct implications on the choice of energies used in the thermal occupation functions of the reservoirs for finite detuning between the optical modes and the two-level system.

In this work, we stay within the realm of the quantum master Eq.~\eqref{EqMasterEquation} with Lindblad-type Liouvillians $\mathcal{L}_\alpha$, see Eq.~\eqref{EqLiouvillian},
which is frequently applied to optical systems, see e.g. Ref.~\cite{RestrepoPRL2014}. For the case of classical fields with a time-dependent Hamiltonian $\hat{H}_S(t)$,
thermodynamic aspects of more detailed approaches have been studied in \cite{GevaJChemPhys1996,SzczygielskiPRE2013,ElouardNJP2020}, where also the impact of the Rabi splitting  for strong classical fields is addressed. It should be noted that the derivation of quantum master equations in the presence of time-dependent Hamiltonians $\hat{H}_S(t)$ is a matter of ongoing discussion, see, e.g. \cite{HotzPRA2021} and references cited therein.

\section{Two-level system with electronic reservoirs}\label{Sec2Level}
We consider an electronic two-level system with energies $E_u>E_l$
as given by the bare  Hamiltonian
\begin{equation}
  \hat{H}_0=E_u\hat{c}_u^\dagger\hat{c}_u+E_l\hat{c}_l^\dagger\hat{c}_l
  \label{EqHam0}
\end{equation}
which is coupled to electron reservoirs $\alpha\in\{u,l\}$ with temperatures
 $T_\alpha$ and electrochemical potential $\mu_\alpha$ via Lindblad operators\cite{LindbladCMP1976,BreuerBook2006}
\begin{equation}\begin{split}
\mathcal{L}_\alpha[\hat{\rho}] =& \gamma_\alpha f_\alpha \mathcal{D}_{\hat{c}_\alpha^\dagger}[\hat{\rho}]+\gamma_\alpha (1-f_\alpha)\mathcal{D}_{\hat{c}_\alpha}[\hat{\rho}]\\
\textrm{with }\mathcal{D}_\sigma[\hat{\rho}] =& \sigma\hat{\rho}\sigma^\dagger-
\frac{1}{2}\left(\sigma^\dagger\sigma\hat{\rho}+\hat{\rho}\sigma^\dagger\sigma\right)
\label{EqLiouvillian}
\end{split}\end{equation} 
where $\gamma_\alpha$ denotes the coupling strength between the level $\alpha$ and its connected reservoir.
This is known as the local approach \cite{LevyEPL2014,HoferNJP2017,GonzalezOpenSystInfoDyn2017}. Commonly, one assumes that only the
 energy $E_\alpha$ of the isolated level is relevant for the transition to bath $\alpha$, which results in  the occupation functions $f_\alpha=f_\alpha^\textrm{common}$ with the Fermi function
\begin{equation}
f_\alpha^\textrm{common}=\frac{1}{\exp\left[(E_\alpha-\mu_\alpha)/k_BT_\alpha\right]+1}\, .
\label{EqFguess}
\end{equation}
However, it is known that such an approach can lead to violations of the thermodynamic rules \cite{BoukobzaPRL2007,LevyEPL2014,StockburgerFortschrittePhysik2017}, which is also the case for the system studied here (see below). 
The key point of this work is to trace such violations to the use of Eq.~\eqref{EqFguess}. Therefore,  
we keep the occupations $f_\alpha$ undefined until we can identify effective energies (\ref{EqEffectiveEnergies},\ref{EqEffectiveEnergiesQuant}) which are suggested to replace the localized  level energies $E_\alpha$ in Eq.~\eqref{EqFguess}. 
These effective energies reflect the coupling of the local levels to the other levels in the system (here by the light field), which is disregarded in the common use of the local approach.

Transitions between the states $u$ and $l$ are possible by coupling to an optical field resulting in a net rate $R$ for transitions $u\to l$.
Below, we provide detailed calculations of $R$ in the steady state (superscript$^\textrm{ss}$) both for classical and quantum fields in Sects.
\ref{SecTwoLevelClassical} and \ref{SecTwoLevelQuantized}, respectively.
In particular, we investigate the sign of $R^\textrm{ss}$ obtained by quantum kinetics and compare it with the sign
determined by thermodynamic considerations as outlined in Sec.~\ref{SecThermo}.

Typically, such systems are treated under resonance ($\Delta=0$) in the literature (see, e.g., \cite{BergenfeldtPRL2014,NiedenzuQuantum2019}). Here the focus is on detuning with a finite value of $\Delta$. 

\subsection{Classical field}\label{SecTwoLevelClassical}
Within the common rotating wave approximation (RWA) we set $\hat{H}_S=\hat{H}_0+\hat{V}_\textrm{cl}(t)$ with
\begin{equation}
\hat{V}_\textrm{cl}(t)=\hbar\epsilon \hat{c}_u^\dagger \hat{c}_l \e^{-\imai\omega t} +\hbar\epsilon^*\hat{c}_l^\dagger\hat{c}_u\e^{\imai\omega t}\label{EqVoft}
\end{equation}
with  the coupling strength $\epsilon$.
A standard density matrix calculation, as detailed in Sec.~\ref{SecDetailClassics}, provides the following results in the steady state,
where $R^\textrm{ss}$ is the transition rate between the upper and lower level, as given by Eq.~\eqref{EqRssFinal}:
The net power absorbed from the electromagnetic field \eqref{EqWQquantumFull} is given by
\begin{equation}
  P_s^\textrm{ss}=-\hbar\omega R^\textrm{ss}
  \label{EqPowerSSClassical}
\end{equation}
and the particle and energy flows from the contacts \eqref{EqEnergyLeadAlpha} in the
steady state read
\begin{equation}
  \dot{N}^\textrm{ss}_{u}= R^\textrm{ss} \qquad \textrm{and }
\dot{N}^\textrm{ss}_{l}= -R^\textrm{ss}
\label{EqParticleSSClassical}
\end{equation}
\begin{equation}
\dot{U}^\textrm{ss}_{u}= \tilde{E}_{u} R^\textrm{ss} \qquad \textrm{and }
\dot{U}^\textrm{ss}_{l}= -\tilde{E}_{l} R^\textrm{ss}
\label{EqEnergySSClassical}
\end{equation}
with the effective energies
\begin{equation}
\tilde{E}_{u}=E_u+\frac{\gamma_u \hbar\Delta }{\gamma_u+\gamma_l}\qquad\textrm{and }
\tilde{E}_{l}=E_l-\frac{\gamma_l \hbar\Delta }{\gamma_u+\gamma_l}\, .
\label{EqEffectiveEnergies}
\end{equation}
Note that this is a rigorous result for the system considered based on the general thermodynamic framework of section \ref{SecThermo} within the Markovian approximation and the local coupling to the reservoir \eqref{EqLiouvillian}. The effective energies can be interpreted as a distribution of the total detuning $\hbar\Delta$ to the bare levels according to their relative coupling strengths $\gamma_\alpha$, so that 
they satisfy $\tilde{E}_{u}-\tilde{E}_{l}=\hbar\omega$. This reflects the fact that the sum of energy flows, $\dot{U}^\textrm{ss}_{u}+\dot{U}^\textrm{ss}_{l}+P_S=0$
conserves energy \eqref{EqEnergySteadyState}. 

Based on these results the entropy production \eqref{EqEntropyLeadAlpha} provides
\begin{equation}
\dot{S}^\textrm{ss}_u+\dot{S}^\textrm{ss}_l=R^\textrm{ss}\left(\frac{\tilde{E}_{l}-\mu_l}{T_l}-\frac{\tilde{E}_{u}-\mu_u}{T_u}\right)\, .\label{EqSclassical}
\end{equation}
Up to now, we performed all  calculations without specifying the vales of $f_\alpha$, which need to reflect the reservoir properties. Now we use the criterion of  positive definite
entropy production \eqref{EqEntropySteadyState} to identify these. According to Eq.~\eqref{EqRssFinal},
$R^\textrm{ss}$ has the same sign as $(f_u-f_l)$. Thus, positivity of entropy production \eqref{EqEntropySteadyState}  implies that $f_u-f_l$ needs to have the same sign as 
$\left(\frac{\tilde{E}_{l}-\mu_l}{T_l}-\frac{\tilde{E}_{u}-\mu_u}{T_u}\right)$. This 
can be guaranteed if we choose
\begin{equation}
f_\alpha=\frac{1}{\e^{(\tilde{E}_{\alpha}-\mu_\alpha)/k_BT_\alpha}+1}
\label{EqfwithEtilde}
\end{equation}
as $\left(\frac{1}{\e^{X_u}+1}-\frac{1}{\e^{X_l}+1}\right)$ has always the same sign as
$(X_l-X_u)$ due to the strong monotonic decrease of $\frac{1}{\e^X+1}$. However, violations of Eq.~\eqref{EqEntropySteadyState} are possible if we choose the  bare energy levels $E_\alpha$ in the bath Fermi functions following the tempting guess (\ref{EqFguess}).
We conclude, that the effective energies \eqref{EqEffectiveEnergies} should be used for the occupation functions in the fermionic reservoirs in full analogy to the bosonic case treated in \cite{KalaeePRA2021}.

For equal temperatures, $T_u=T_l=T$, Eq.~\eqref{EqSclassical} shows that $R^\textrm{ss}$ has the same sign as $\mu_u-\mu_l-\hbar\omega$. This implies operation as an LED if the bias  $\mu_u-\mu_l$ surpasses the photon energy $\hbar\omega$ and operation as a solar cell in the opposite case. This is a consequence of the fact that no entropy is produced in the light field, which is described by a classical field in Eq.~\eqref{EqVoft}. Thus, the only source of entropy production
is the generation of heat from the excess  energy $\mu_u-\mu_l>\hbar\omega$ for emission or $\hbar\omega >\mu_u-\mu_l>$ for absorption, which is transferred to the reservoirs. 

Note, that this procedure also applies to tunneling problems, which correspond to $\omega=0$. Thereby it resolves the violation of the second law for the local approach in Ref.~\cite{LevyEPL2014}.

\subsection{Quantized field}\label{SecTwoLevelQuantized}
Within the RWA we set in the spirit of the Jaynes-Cummings model \cite{JaynesProcIEEE1963}
\[ 
\hat{V}_\textrm{JC}=\hbar g \hat{c}_u^\dagger \hat{c}_l \hat{a} +\hbar g^*\hat{c}_l^\dagger\hat{c}_u\hat{a}^\dagger
+\hbar \omega_\textrm{cav} \hat{a}^\dagger\hat{a}
\]
with the bosonic annihilation operator $\hat{a}$ for the photon mode. Now the detuning is given by $\hbar\Delta_\textrm{cav}=\hbar \omega_\textrm{cav}+E_l-E_u$.
In order to allow for a steady state we add the interaction of the photon mode with a thermal bosonic reservoir with average occupation $n_\textrm{b}$ and transition rate $\gamma_\textrm{b}$ via the Lindblad operator
\begin{equation}
\mathcal{L}_\textrm{b}[\hat{\rho}]= 
\gamma_\textrm{b} (n_\textrm{b}+1)\mathcal{D}_{\hat{a}}[\hat{\rho}]
+\gamma_\textrm{b} n_\textrm{b}\mathcal{D}_{\hat{a}^\dagger}[\hat{\rho}]\, .
\end{equation}
As for the fermionic occupation factors $f_\alpha$, the value of $n_b$ is specified at a later stage.
This provides the  equation of motion for the density operator
\begin{equation}
\frac{\d \hat{\rho}}{\d t}=
\frac{1}{\imai\hbar} [\hat{H}_0+\hat{V}_\textrm{JC},\hat{\rho}] +\mathcal{L}_u[\hat{\rho}]+\mathcal{L}_l[\hat{\rho}]+\mathcal{L}_\textrm{b}[\hat{\rho}]
\label{EqMasterJC}
\end{equation}
As described in Appendix \ref{SecDetailQuant}, this provides the following results in the steady state
with the net transition rate $R^\textrm{ss}$, for which we do not have a closed expression:
\begin{equation}
  \dot{N}^\textrm{ss}_{u}=R^\textrm{ss} \qquad \textrm{and } 
  \dot{N}^\textrm{ss}_{l}= -R^\textrm{ss}\, , \label{EqParticleSSQuant}
\end{equation}
  \begin{equation}\begin{split}
  \dot{U}^\textrm{ss}_{u}= \tilde{E}_{u} R^\textrm{ss}\, , \qquad
  \dot{U}^\textrm{ss}_{l}=& - \tilde{E}_{l} R^\textrm{ss}\, ,  \\
\textrm{and } 
  \dot{U}^\textrm{ss}_\textrm{b}=& - \tilde{E}_\textrm{ph} R^\textrm{ss}\label{EqEflowQuant}
\end{split}\end{equation}
with the effective energies
\begin{equation}\begin{split}
\tilde{E}_{u}=&E_u+\frac{\gamma_u \hbar\Delta_\textrm{cav} }{\gamma_u+\gamma_l+\gamma_\textrm{b}}\, , 
\\
\tilde{E}_{l}=&E_l- \frac{\gamma_l \hbar\Delta_\textrm{cav}}{\gamma_u+\gamma_l+\gamma_\textrm{b}}\, , 
\\ 
\textrm{and }
\tilde{E}_\textrm{ph}=&\hbar\omega_\textrm{cav}-\frac{\gamma_\textrm{b} \hbar\Delta_\textrm{cav} }{\gamma_u+\gamma_l+\gamma_\textrm{b}}
\, ,
\label{EqEffectiveEnergiesQuant}
\end{split}\end{equation}
which satisfy $\tilde{E}_{u}=\tilde{E}_{l}+\tilde{E}_\textrm{ph}$, so that the fluxes \eqref{EqEflowQuant} fulfil energy conservation
(\ref{EqEnergySteadyState}). This is analogous to Eq.~\eqref{EqEffectiveEnergies} for the classical treatment. However, the finite lifetime of the photon mode in the cavity 
contributes now to the broadening and, correspondingly, we obtain an effective photon energy $\tilde{E}_\textrm{ph}$, which is relevant for the transitions with the bosonic bath.

Now we consider the requirement for positive entropy production \eqref{EqEntropySteadyState}. 
Applying the energy and particle currents above and  Eq.~\eqref{EqEntropyLeadAlpha}, the entropy condition reads
\begin{equation}
R^\textrm{ss}\left[
\frac{\tilde{E}_\textrm{ph}}{T_\textrm{b}}+\frac{\tilde{E}_l-\mu_l}{T_l}
-\frac{\tilde{E}_u-\mu_u}{T_u}\right]\ge 0\,
\label{EqEntropyCondQuant}.
\end{equation} 
where $T_\textrm{b}$ is the temperature of the bosonic bath coupled to the photon mode.

\begin{widetext}
In Appendix \ref{SecDetailQuant} it is shown within a Hartree-Fock like approximation, that 
\begin{equation}
\textrm{The sign of }R^\textrm{ss}\textrm{ equals the sign of }
\frac{f_u}{1-f_u} -
\frac{f_l}{1-f_l}\times \frac{n_\textrm{b}}{1+n_\textrm{b}}
\label{EqSignQuantum}
\end{equation}
Applying the Fermi and Bose distributions 
\[
f_u=\frac{1}{\e^{(\tilde{E}_u-\mu_u)/k_BT_u}+1}\, , \qquad
f_l=\frac{1}{\e^{(\tilde{E}_l-\mu_l)/k_BT_l}+1}\, , \qquad \textrm{and }
n_\textrm{ph}=\frac{1}{\e^{\tilde{E}_\textrm{ph}/k_BT_\textrm{b}}-1}\, ,
\]
Eq.~\eqref{EqSignQuantum} becomes
\[
\e^{-(\tilde{E}_u-\mu_u)/k_BT_u}-\e^{-\tilde{E}_\textrm{ph}/k_BT_\textrm{b}}\times\e^{-(\tilde{E}_l-\mu_l)/k_\textrm{b}T_l}
\quad\textrm{determines the sign of }R^\textrm{ss}
\]
which guaranties the positivity in Eq.~\eqref{EqEntropyCondQuant} due to the
monotonic increase of the exponential function.
Here it is crucial, that the effective energies \eqref{EqEffectiveEnergiesQuant} enter the respective occupation functions. Otherwise violations of Eq.~\eqref{EqEntropyCondQuant} can be easily constructed for particular values of temperatures and electrochemical potentials.
\end{widetext}

For equal temperatures, $T_u=T_l=T$, Eq.~\eqref{EqEntropyCondQuant}
shows that $R^\textrm{ss}$ has the same sign as $\mu_u-\mu_l-\tilde{E}_\textrm{ph}(1-T/T_\textrm{b})$. Thus, light emission ($R^\textrm{ss}>0$) is even possible for
$\mu_u-\mu_l<\tilde{E}_\textrm{ph}$ provided that $T_\textrm{b}$ is not too large.
In this case the LED emits more light than it consumes electrical energy, which has been experimentally verified \cite{SanthanamPRL2012}. The excess energy is taken from the reservoirs which provides cooling \cite{SadiNatPhotonics2020}. For operation as a solar cell ($R^\textrm{ss}<0$) this shows that the extracted electrical power $P_\textrm{el}=-R^\textrm{ss}(\mu_u-\mu_l)$ is limited by
\[
P_\textrm{el}\le \dot{U}_\textrm{b}^\textrm{ss}\frac{T_\textrm{b}-T}{T_\textrm{b}}
\]
which is just Carnot's law for the incoming heat
$\dot{U}_\textrm{b}^\textrm{ss}$ from the warm reservoir with $T_\textrm{b}>T$. 

\section{Relation to frequency pulling in a laser}\label{SecTwoLevelSemiclassical}
For laser operation, the coherent states are a better approximation for the optical field than number states \cite{MandelBook1995}. Technically,  this can be done by the semi-classical approximation \cite{ShirleyPR1969}, where $\langle \hat{a}\rangle$ is treated as a classical field. Therefore, in Eqs.~(\ref{EqSigmaUUquant},\ref{EqSigmaLLquant}) the approximation
\begin{equation}
  Y=g^*\textrm{Tr}\{\hat{c}_l^\dagger\hat{c}_u \hat{a}^\dagger \hat{\rho}\}\approx g^* a^*\sigma_{ul}\label{EqSemiclassical}
\end{equation}
is used with
\[
\sigma_{ul}=\e^{\imai \omega t} \textrm{Tr}\{\hat{c}_l^\dagger\hat{c}_u\hat{\rho}\}\quad\textrm{and } a=\e^{\imai \omega t}\textrm{Tr}\{\hat{a} \hat{\rho}\}
\]
where a natural oscillation with frequency $\omega$ is assumed. Without detuning, $\omega=\omega_\textrm{cav}=(E_u-E_l)/\hbar$ is the natural choice,  which corresponds to the interaction picture.  
For the case of detuning, $\omega$ is determined below. Eq.~\eqref{EqMasterJC} provides
\begin{eqnarray}
\frac{\d}{\d t} \sigma_{ul}&=&\imai \left(\omega-\frac{E_u-E_l}{\hbar}\right) \sigma_{ul}+\imai g^*a^*  (\sigma_{uu}-\sigma_{ll})
\nonumber \\
&&-\frac{\gamma_u+\gamma_l}{2}\sigma_{ul}
\label{EqSigmaULSemiclassical}\\
\frac{\d}{\d t}  a&=&\imai \left(\omega-\omega_\textrm{cav}\right) a-\imai g^*\sigma_{ul} -\frac{\gamma_\textrm{b}}{2}  a
\label{EqASemiclassical}
\end{eqnarray}
which,   together with (\ref{EqSigmaUUquant},\ref{EqSigmaLLquant}), give a closed set of equations.
In the steady state Eq.~\eqref{EqASemiclassical} provides
\begin{equation}
\sigma_{ul}^\textrm{ss}=\frac{\omega-\omega_\textrm{cav}+\imai  \gamma_\textrm{b}/2}{|g|^2}g a^\textrm{ss}\label{EqrhoULSSSemiclassical1}
\end{equation}
With the semiclassical approximation \eqref{EqSemiclassical}, the set of
equations for the quantum case
(\ref{EqSigmaUUquant},\ref{EqSigmaLLquant},\ref{EqSigmaULSemiclassical}) equal the classical treatment
(\ref{EqSigmaUU},\ref{EqSigmaLL},\ref{EqSigmaUL}) with
$\epsilon=ga$. Thus we obtain from Eqs.~(\ref{EqrhoULSS},
\ref{EqRelSigmaF})
\begin{equation}
  \sigma^\textrm{ss}_{ul}=\frac{- H(|g a^\textrm{ss}|^2)
    (f_u-f_l)}{\Delta+\imai(\gamma_u+\gamma_l)/2}ga^\textrm{ss}
\label{EqrhoULSSSemiclassical2}
\end{equation}
Eqs.~(\ref{EqrhoULSSSemiclassical1},\ref{EqrhoULSSSemiclassical2}) are
only consistent, if
\begin{multline}
(\omega-\omega_\textrm{cav}+\imai  \gamma_\textrm{b}/2) [\Delta+
    \imai(\gamma_u+\gamma_l)/2]
\\
=-|g|^2H(|ga^\textrm{ss}|^2)
  (f_u-f_l)\label{EqConsistency}
  \end{multline}
While, the real part of the equation  provides the field strength
$a^\textrm{ss}$ of laser light in the cavity due to gain saturation,
its imaginary part determines  $\omega$: As the right-hand side is
purely real, the imaginary part needs to be zero on the left-hand
side. Using $\Delta=\omega-(E_u-E_l)/\hbar$, one obtains
\begin{equation}
  \hbar\omega=\frac{(\gamma_u+\gamma_l)\hbar\omega_\textrm{cav}+\gamma_\textrm{b}(E_u-E_l)}{\gamma_u+\gamma_l+\gamma_\textrm{b}}\,
  .\label{EqFrequencyPull}
\end{equation}
Straightforward algebra shows, that
$\hbar\omega=\tilde{E}_\textrm{ph}$ from
Eq.~\eqref{EqEffectiveEnergiesQuant}. Thus the effective photon energy
obtained in Sec.~\ref{SecTwoLevelQuantized} matches the actual
oscillation frequency in the laser. Furthermore, the effective energies
$\tilde{E}_u$ and $\tilde{E}_l$ from the classical
field~\eqref{EqEffectiveEnergies} and the quantum
treatment~\eqref{EqEffectiveEnergiesQuant} agree with each other for
this choice of $\hbar\omega$.

Eq.~\eqref{EqFrequencyPull} shows that for detuning between the cavity
frequency and the optical transition frequency, the laser operates at
a frequency in between. This is known as \textit{frequency pulling}
\cite{SiegmanBook1986}. Actually, the expression (12.23) of
\cite{SiegmanBook1986} is directly obtained from Eq.~\eqref{EqFrequencyPull} by introducing the
Q-factors $Q_c=\omega/\gamma_\textrm{b}$ and
$Q_a=\omega/(\gamma_u+\gamma_l)$ for the cavity and the atomic
transition, respectively. Thus, the treatment by heat flows suggested
here,  provides the same result as a detailed optoelectronic study.


\section{Relation to Bloch gain}\label{SecBloch}
For layered semiconductor structures, optical transitions occur
between subbands with a well-defined transition energy $E_u-E_l$ due
the quantized energies in growth direction \cite{HelmReview1999}. In
addition, the lateral free-particle motion (with energy $E_k$
described by wave vector $k$) provides a continuous degree of freedom
with  occupation function $f_u(E_k)$ and $f_l(E_k)$ in each of the
subbands. Due to intra-subband scattering, the levels are broadened and
thus light emission/absorption is possible for detuning, with a finite
value of $\Delta= \omega-(E_u-E_l)/\hbar$. A microscocopic density-matrix approach for the scattering \cite{WillenbergPRB2003} provided
the steady state net transition rate between the upper and the lower
subband 
\begin{widetext}
\begin{equation}
R^\textrm{ss}_\textrm{DM}(k_0)\propto 
\frac{\gamma_u+\gamma_l}{\Delta^2+(\gamma_u+\gamma_l)^2/4}
\left\{\frac{\gamma_l\left[ f_u(E_{k_0})-f_l(E_{k_0}-\hbar\Delta)\right]}{\gamma_u+\gamma_l}+\frac{\gamma_u\left[f_u(E_{k_0}+\hbar\Delta)-f_l(E_{k_0})\right]}{\gamma_u+\gamma_l}\right\}\label{EqWillenberg}
\end{equation}
\end{widetext}
for states with a particular value of $k=k_0$ (which is essentially
conserved in the optical transition), see Eq.~(20) or
Ref.~\cite{WillenbergPRB2003}.  The first factor provides the common
Lorentzian broadening of the line and the second factor shows, that
the transitions are driven by differences between the occupations of
the subbands. This factor is not just  $f_u(E_{k_0})-f_l(E_{k_0})$ as
frequently assumed, but has different energy arguments. This leads to
a particular gain spectrum for the case of equal occupation of both
subbands $[f_u(E)=f_l(E)]$, called dispersive gain or Bloch gain due
to its relation to Bloch oscillations in
superlattices\cite{KtitorovSovPhysSolState1972,WackerPhysRep2002}. This
type of gain could be observed in Quantum Cascade Lasers
\cite{TerazziNatPhys2007} and has been recently suggested to be
relevant for the generation of frequency combs \cite{OpacakPRL2021}.

Analysing the occupation factors of Eq.~\eqref{EqWillenberg} reveals
that the occupations of the upper level are taken at an average energy
$\tilde{E}^u_{k_0}=
E_{k_0}+\hbar\Delta\gamma_u/(\gamma_u+\gamma_l)$. Correspondingly, the
occupations of the lower subband are taken at an average energy
$\tilde{E}^l_{k_0}= E_{k_0}-\hbar\Delta\gamma_l/(\gamma_u+\gamma_l)$,
so that
$E_u+\tilde{E}^u_{k_0}-(E_l+\tilde{E}^l_{k_0})=\hbar\omega$. This
fully corresponds to the choice \eqref{EqEffectiveEnergies} obtained
above, which therefore provides the same characteristic Bloch gain.
The average energies \eqref{EqEffectiveEnergies} can alternatively be
obtained as average transition energies within a
Green's function treatment \cite{KalaeePRA2021} (see
Ref.~\cite{WackerNatPhys2007} for the connection with Bloch gain).


\section{Conclusion}
Applying a quantum thermodynamic approach, effective energies were
identified to describe optical transition under detuning, as given in
Eqs.~(\ref{EqEffectiveEnergies},\ref{EqEffectiveEnergiesQuant}) for
the classical and quantum treatment, respectively. These effective
energies satisfy energy conservation by distributing the detuning to
the bare energies according to their respective contribution to the
broadening of the transition. Applying these effective energies for
transitions with thermal reservoirs provides thermodynamic consistency within the generally accepted heat and work definitions (\ref{EqEnergyLeadAlpha}\ref{EqWQquantumFull}), where both the first and second law are
satisfied for steady state operation.

The effective energy of the photon could be attributed to the pulled
frequency in a laser cavity. Similarly, the effective level energies
agree with the average energies associated with inter-subband
transitions. Thus, the approach used here provides a new comprehensive
view of features obtained from different sophisticated microscopic
treatments.

All results obtained are rigorous within the Markovian Lindblad master equation used. The only exception is
 the fulfillment of the second law for the quantum treatment of the cavity mode, where a Hartree-Fock like
approximation was required. It would be interesting, how far the results can be generalized. Another open issue is whether these effective energies can also be applied for noise calculations, where different types of averages apply.

It is interesting to note, that the effective energies appear independently whether
the coupling to the electromagnetic field is treated classically by a time-dependent Hamiltonian and quantum-mechanically within a time-independent Hamiltonian. Thus, these findings appear not 
to be affected by the question, in how far the quantum master equation Eq.~\eqref{EqMasterEquation} is applicable for time-dependent Hamiltonians. 

For the classical case, the results can be generalized to arbitrary fermionic systems with a time-periodic Hamiltonian, see Appendix~\ref{SecGeneralFermion}. However, the identification of the effective energies relies then on a self-consistent solution of the system dynamics with the occupation functions, which may limit practical applications. 
(The recently developed thermodynamically consistent local approach\cite{PottsNJP2021} is an interesting alternative at the cost of a limited resolution for heat.)
Albeit the approach is thermodynamic consistent for arbitrary system parameters, one has to remember that the local approach \eqref{EqLiouvillian} restricts to a single transition energy, which does not allow the detailed description of photon-assisted reservoir transitions or Coulomb blockade phenomena.


\section{Acknowledgment}
The author thanks Alex Kalaee, Patrick Potts, and Peter Samuelsson for helpful discussions and collaboration on related issues.
Financial support from the Knut and Alice Wallenberg Foundation
(project 2016.0089), the Swedish Research Council (project
2017-04287), and NanoLund is gratefully acknowledged.

\appendix


\section{Calculations for the classical case}\label{SecDetailClassics}
The quantum master equation \eqref{EqMasterEquation} with $\hat{H}_S=\hat{H}_0+\hat{V}_\textrm{cl}(t)$ based on the operators (\ref{EqHam0},\ref{EqVoft})
can be mapped to a time-independent problem in the rotating frame where
$\hat{A}^R=\hat{U}(t)\hat{A}\hat{U}^\dagger(t)$ with $\hat{U}(t)=
\e^{\imai[(E_l+\hbar\omega)\hat{c}_u^\dagger\hat{c}_u+ E_l \hat{c}_l^\dagger\hat{c}_l]t/\hbar}$ and provides the equation of motion for the density operator
\[\begin{split}
\frac{\d \hat{\rho}^R}{\d t}=&
-\imai [-\Delta \hat{c}_u^\dagger\hat{c}_u+\epsilon \hat{c}_u^\dagger\hat{c}_l  +\epsilon^*\hat{c}_l^\dagger\hat{c}_u ,\hat{\rho}^R ] \\
&+\mathcal{L}_u[\hat{\rho}^R]+\mathcal{L}_l[\hat{\rho}^R]
\end{split}\]
with $\Delta=\omega-(E_u-E_l)/\hbar$. This gives the
equations of motion for the reduced density matrix
$\sigma_{ij}=\textrm{Tr}\{\hat{c}_j^\dagger\hat{c}_i \hat{\rho}^\textrm{R}\}$

\begin{eqnarray}
\frac{\d}{\d t} \sigma_{uu}&=&\gamma_u (f_u- \sigma_{uu})
+\imai(\epsilon^* \sigma_{ul}-\epsilon\sigma_{lu})\label{EqSigmaUU}\\
\frac{\d}{\d t} \sigma_{ll}&=&\gamma_l( f_l- \sigma_{ll})
+\imai(\epsilon \sigma_{lu}-\epsilon^*\sigma_{ul})\label{EqSigmaLL}\\
\frac{\d}{\d t} \sigma_{ul}&=&\imai \Delta \sigma_{ul}+\imai\epsilon(\sigma_{uu}-\sigma_{ll})-\frac{\gamma_u+\gamma_l}{2}\sigma_{ul}
\label{EqSigmaUL}
\end{eqnarray}

From the changes of the populations, we identify the net transitions rate $u\to l$
\begin{equation}
  R= -\imai(\epsilon^*\sigma_{ul}-\epsilon\sigma_{ul}^*)=2\Im\{\epsilon^*\sigma_{ul}\}\, .
\label{EqRClassical}
\end{equation}
Furthermore the work flow \eqref{EqWQquantumFull} becomes
\begin{equation}\begin{split}
P_S=& \textrm{Tr}\left\{\imai\hbar\omega\left(\epsilon^*\hat{c}_l^\dagger\hat{c}_u\e^{\imai\omega t}-\epsilon \hat{c}_u^\dagger \hat{c}_l \e^{-\imai\omega t}\right) \hat{\rho}\right\}\\
=&\imai\hbar\omega\textrm{Tr}\left\{\left(\epsilon^*\hat{c}_l^\dagger\hat{c}_u- \epsilon\hat{c}_u^\dagger \hat{c}_l \right) \hat{\rho}^R\right\}=-\hbar \omega R\, .\label{EqPowerClassical}
\end{split}\end{equation}
This shows that each transition from the upper to the lower level is associated with the energy portion $\hbar\omega$ removed from the system even if $E_u-E_l\neq \hbar\omega$. This agrees with the conception that the transition is associated with the creation of one photon with angular frequency $\omega$ albeit the optical field in treated as a classical variable in Eq.~\eqref{EqVoft}. In the steady state, Eq.~\eqref{EqPowerClassical} directly provides Eq.~\eqref{EqPowerSSClassical}.

The  energy and particle flows from the leads \eqref{EqEnergyLeadAlpha} are 
\begin{equation}\begin{split}
\dot{U}_\alpha=& \textrm{Tr}\{\hat{H}\mathcal{L}_\alpha[\hat{\rho}]\}=
\textrm{Tr}\{\hat{H}^R\mathcal{L}_\alpha[\hat{\rho}^R]\}
\\
=&E_\alpha\gamma_\alpha (f_\alpha- \sigma_{\alpha\alpha})-\frac{\hbar\gamma_\alpha}{2}(\epsilon\sigma_{lu}+\epsilon^*\sigma_{ul})
\label{EqEnergyClassical}
\end{split}\end{equation}
and
\begin{equation}\begin{split}
\dot{N}_\alpha=& \textrm{Tr}\{(\hat{c}^\dagger_u\hat{c}_u+ \hat{c}^\dagger_l\hat{c}_l)\mathcal{L}_\alpha[\hat{\rho}]\}=
\gamma_\alpha (f_\alpha- \sigma_{\alpha\alpha})\, .
\label{EqParticleClassical}
\end{split}\end{equation}

In the steady state, Eq.~\eqref{EqSigmaUL} provides
\begin{equation}
\sigma^\textrm{ss}_{ul}=\frac{-\epsilon(\sigma_{uu}-\sigma_{ll})}{\Delta+\imai(\gamma_u+\gamma_l)/2}
\label{EqrhoULSS}
\end{equation}
Thus, we can relate the transition rate \eqref{EqRClassical} to the real part of $\epsilon^*\sigma^\textrm{ss}_{ul}$ by
 \begin{equation}
\Re\{\epsilon^*\sigma_{ul}^\textrm{ss}\}=-\frac{2\Delta}{\gamma_u+\gamma_l} \Im\{\epsilon^*\sigma_{ul}^\textrm{ss}\}=-\frac{\Delta}{\gamma_u+\gamma_l}R^\textrm{ss}
\label{EqReSigma}
\end{equation}
In the steady state, Eqs.~(\ref{EqSigmaUU},\ref{EqSigmaLL}) become
\begin{equation}
\gamma_u (f_u- \sigma_{uu})=R^\textrm{ss}\qquad\textrm{and } \gamma_l (f_l- \sigma_{ll})=-R^\textrm{ss} \label{EqFlows} 
\end{equation}
and thus Eq.~\eqref{EqParticleClassical} provides Eq.~(\ref{EqParticleSSClassical}) of the main text.
Inserting Eqs.~(\ref{EqReSigma},\ref{EqFlows}) into the energy flows \eqref{EqEnergyClassical} from the leads provides
Eqs.~(\ref{EqEnergySSClassical},\ref{EqEffectiveEnergies}), which are the key result.

Eqs.~(\ref{EqRClassical},\ref{EqrhoULSS}) provide the steady state transition rate 
\[
R^\textrm{ss}=\alpha (\sigma_{uu}-\sigma_{ll})\quad \textrm{with }
\alpha=\frac{|\epsilon|^2(\gamma_u+\gamma_l)}{(\gamma_u+\gamma_l)^2/4+\Delta^2}
\]
Then Eqs.~(\ref{EqSigmaUU},\ref{EqSigmaLL}) result in
\begin{widetext}
\[
\begin{pmatrix}
  \gamma_u+\alpha & -\alpha \\
  -\alpha & \gamma_l+\alpha
\end{pmatrix}
\begin{pmatrix}
  \sigma_{uu}^\textrm{ss}\\
  \sigma_{ll}^\textrm{ss}
\end{pmatrix}
    =
    \begin{pmatrix}\gamma_u f_u\\
      \gamma_l f_l
\end{pmatrix}
\Rightarrow
\begin{pmatrix}
  \sigma_{uu}^\textrm{ss}\\
  \sigma_{ll}^\textrm{ss}
\end{pmatrix}
=\frac{1}{\gamma_u\gamma_l+\alpha(\gamma_u+\gamma_l)}
\begin{pmatrix}\alpha(\gamma_u f_u+ \gamma_l f_l)+\gamma_u\gamma_l f_u\\
\alpha(\gamma_u f_u+ \gamma_l f_l)+\gamma_u\gamma_l f_l
\end{pmatrix}
\]
so that
\begin{equation}
\sigma_{uu}^\textrm{ss}-\sigma_{ll}^\textrm{ss}=H(|\epsilon|^2)(f_u-f_l)\quad
\textrm{with }H(|\epsilon|^2)=\frac{\gamma_u\gamma_l}{\gamma_u\gamma_l+\alpha(\gamma_u+\gamma_l)}
=\frac{1}{1+|\epsilon|^2\frac{(\gamma_u+\gamma_l)^2}{\gamma_u\gamma_l[(\gamma_u+\gamma_l)^2/4+\Delta^2]}}
\label{EqRelSigmaF}
\end{equation}
\end{widetext}
and we find
\begin{equation}
  R^\textrm{ss}=\alpha H(f_u-f_l)\label{EqRssFinal}
\end{equation}
with positive $\alpha$ and $H$.


\section{Calculations the quantum case}\label{SecDetailQuant}

Defining the average occupation of the photon mode $n_\textrm{ph}=\textrm{Tr}\{\hat{a}^\dagger\hat{a} \hat{\rho}\}$ and  $Y=g^*\textrm{Tr}\{\hat{c}_l^\dagger\hat{c}_u \hat{a}^\dagger \hat{\rho}\}$, Eq.~\eqref{EqMasterJC} provides the equations of motion
\begin{eqnarray}
\frac{\d}{\d t} \sigma_{uu}&=&\gamma_u (f_u- \sigma_{uu})
+\imai (Y-Y^*)\label{EqSigmaUUquant}\\
\frac{\d}{\d t} \sigma_{ll}&=&\gamma_l (f_l- \sigma_{ll})
-\imai(Y-Y^*)\label{EqSigmaLLquant}\\
\frac{\d}{\d t} n_\textrm{ph}&=&\gamma_\textrm{b}(n_\textrm{b}-n_\textrm{ph})
-\imai(Y-Y^*)
\label{Eqnoptquant}\\
\frac{\d}{\d t} Y&=&\imai \Delta_\textrm{cav} Y+\imai|g|^2 F
-\frac{\gamma_u+\gamma_l+\gamma_\textrm{b}}{2}Y \label{EqSigmaULquant}\\
\textrm{with }F&=&\textrm{Tr}\left\{\hat{c}_u^\dagger\hat{c}_u (1-\hat{c}_l^\dagger\hat{c}_l )\hat{\rho}\right\}\nonumber\\
&&+\textrm{Tr}\left\{(\hat{c}_u^\dagger\hat{c}_u-\hat{c}_l^\dagger\hat{c}_l )\hat{a}^\dagger\hat{a} \hat{\rho}\right\}
\label{EqDefFQuant}
\end{eqnarray}
where $\Delta_\textrm{cav}=\omega_\textrm{cav}-(E_u-E_l)/\hbar$.
Eqs.~(\ref{EqSigmaUUquant},\ref{EqSigmaLLquant},\ref{Eqnoptquant}) have a very transparent interpretation: The occupations of the levels and the photon mode are fed/emptied by the respective reservoirs with corresponding rates $\gamma_\alpha$, where the speed is proportional to the occupation differences. In addition, there are optical transitions  $u\to l$ generating photons with a rate
\begin{equation}
  R=2\Im\{Y\}\, . \label{EqRateQuant}
\end{equation}
The new quantity $Y$ satisfies a linear differential equation, where the inhomogeneity is determined by the expression $F$ related to  spontaneous [$\propto \textrm{Tr}\{\hat{c}_u^\dagger\hat{c}_u (1-\hat{c}_l^\dagger\hat{c}_l )\hat{\rho}\}$] and stimulated  [$\propto \textrm{Tr}\{(\hat{c}_u^\dagger\hat{c}_u-\hat{c}_l^\dagger\hat{c}_l )\hat{a}^\dagger\hat{a} \hat{\rho}\}$] emission. These provide further dynamical variables, so that we do not have a closed system of equations.

The energy flows from the fermionic reservoirs \eqref{EqEnergyLeadAlpha} are 
\begin{equation}\begin{split}
\dot{U}_\alpha=& \textrm{Tr}\{\hat{H}_S\mathcal{L}_\alpha[\hat{\rho}]\}
\\
=&E_\alpha\gamma_\alpha (f_\alpha- \sigma_{\alpha\alpha})-\frac{\hbar\gamma_\alpha}{2}(Y+Y^*)
\label{EqEnergyAlphaQuant}
\end{split}\end{equation}
and the corresponding energy flow from the bosonic reservoir is
\begin{equation}\begin{split}
\dot{U}_\textrm{b}= &\textrm{Tr}\{\hat{H}_S\mathcal{L}_\textrm{b}[\hat{\rho}]\}
\\
=&\hbar\omega_\textrm{cav} \gamma_\textrm{b} (n_\textrm{b}-n_\textrm{ph})-\frac{\hbar\gamma_\textrm{b}}{2}(Y+Y^*)
\label{EqEnergyBQuant}
\end{split}\end{equation}

Let us now consider the steady state.
Then Eq.~\eqref{EqSigmaULquant} provides
\begin{equation}
Y^\textrm{ss}=-
\frac{|g|^2 F}{\Delta_\textrm{cav}+\imai\frac{\gamma_u+\gamma_l+\gamma_\textrm{b}}{2}}
\,,\label{EqYssQuant}
\end{equation}
which implies
\begin{equation}
\Re\{Y^\textrm{ss}\}=-\frac{2\Delta_\textrm{cav}\Im\{Y^\textrm{ss}\}}{\gamma_u+\gamma_l+\gamma_\textrm{b}}=-\frac{\Delta_\textrm{cav} R^\textrm{ss}}{\gamma_u+\gamma_l+\gamma_\textrm{b}}\, .\label{EqReYss}
\end{equation}
Furthermore, Eqs.~(\ref{EqSigmaUUquant},\ref{EqSigmaLLquant},\ref{Eqnoptquant})  provide
\begin{equation}\begin{split}
\gamma_u (f_u- \sigma_{uu})=R^\textrm{ss}\, , \qquad
\gamma_l (f_l- \sigma_{ll})=&-R^\textrm{ss}\, , 
\\
\quad\textrm{and }\gamma_\textrm{b}(n_\textrm{b}-n_\textrm{ph})=&-R^\textrm{ss}\label{EqRssOcc}
\end{split}\end{equation}
Inserting into Eq.~\eqref{EqParticleClassical} (which is the same here), we obtain
Eq.~\eqref{EqParticleSSQuant}.
Inserting the relations (\ref{EqReYss},\ref{EqRssOcc}) into the  energy flows (\ref{EqEnergyAlphaQuant},\ref{EqEnergyBQuant}), we find in the steady state Eqs.~(\ref{EqEflowQuant},\ref{EqEffectiveEnergiesQuant}).

From Eqs.~(\ref{EqRateQuant},\ref{EqYssQuant}) we get $R^\textrm{ss}=\alpha_Q F$ with
\[
\alpha_Q=\frac{g^2 (\gamma_u+\gamma_l+\gamma_\textrm{b})}{\Delta_\textrm{cav}^2+(\gamma_u+\gamma_l+\gamma_\textrm{b}^2)/4}>0
\]
so that the sign of $R^\textrm{ss}$ equals the sign of $F$.
Using a Hartree-Fock like approximations, Eq.~\eqref{EqDefFQuant} provides
\begin{equation}
  F \approx \sigma_{uu}(1-\sigma_{ll})+(\sigma_{uu}-\sigma_{ll})n_\textrm{ph}\, .
\label{EqHartreeFock}
\end{equation}
Then straightforward algebra shows\begin{widetext}
\begin{equation}
R^\textrm{ss}=\alpha_Q (1-\sigma_{ll})(1-\sigma_{uu}) (1+n_\textrm{ph})
\left( \frac{\sigma_{uu}}{1-\sigma_{uu}}
-\frac{\sigma_{ll}}{1-\sigma_{ll}} \times\frac{n_\textrm{ph}}{1+n_\textrm{ph}}
\right)
\label{EqRssFaktor}
\end{equation}
where we assume $f_u,f_l,\sigma_{uu},\sigma_{ll}<1$ as appropriate for fermionic states, which are not entirely occupied due to their contribution in a dynamical
process. Similarly, Eqs.~\eqref{EqRssOcc} can be rewritten as
\begin{eqnarray}
R^\textrm{ss}&=&\gamma_u (f_u- \sigma_{uu})=\gamma_u (1-\sigma_{uu}) (1-f_u)
\left( \frac{f_u}{1-f_u}- \frac{\sigma_{uu}}{1-\sigma_{uu}}\right)
\label{EqRelfu}
\\
R^\textrm{ss}&=&\gamma_l (\sigma_{ll}-f_l)=\gamma_l(1-\sigma_{ll}) (1-f_l)
\left( \frac{\sigma_{ll}}{1-\sigma_{ll}}- \frac{f_l}{1-f_l}\right)
\label{EqRelfl}\\
R^\textrm{ss}&=&\gamma_\textrm{b}(n_\textrm{ph}-n_\textrm{b})=
\gamma_\textrm{b} (1+n_\textrm{ph})(1+n_\textrm{b})
\left(\frac{n_\textrm{ph}}{1+n_\textrm{ph}}- \frac{n_\textrm{b}}{1+n_\textrm{b}}\right)
\label{EqRelnb}
\end{eqnarray}
Applying Eqs.~(\ref{EqRelfu},\ref{EqRelfl},\ref{EqRelnb}) and subsequently Eq.~\eqref{EqRssFaktor} provides
\[
\begin{array}{l c c c c c c c c}
  R^\textrm{ss}=0 & \Rightarrow
  &\frac{\sigma_{uu}}{1-\sigma_{uu}}=\frac{f_u}{1-f_u} \, ,
  &\frac{\sigma_{ll}}{1-\sigma_{ll}}= \frac{f_l}{1-f_l}
  &\textrm{and} & \frac{n_\textrm{ph}}{1+n_\textrm{ph}}=\frac{n_\textrm{b}}{1+n_\textrm{b}}
  &\Rightarrow & \frac{f_u}{1-f_u} =\frac{f_l}{1-f_l}\times \frac{n_\textrm{b}}{1+n_\textrm{b}}\\
  R^\textrm{ss}>0 & \Rightarrow
  &\frac{\sigma_{uu}}{1-\sigma_{uu}}<\frac{f_u}{1-f_u} \, ,
  &\frac{\sigma_{ll}}{1-\sigma_{ll}}>\frac{f_l}{1-f_l}
  &\textrm{and} & \frac{n_\textrm{ph}}{1+n_\textrm{ph}}>\frac{n_\textrm{b}}{1+n_\textrm{b}}
  &\Rightarrow & \frac{f_u}{1-f_u} >\frac{f_l}{1-f_l}\times \frac{n_\textrm{b}}{1+n_\textrm{b}}\\
  R^\textrm{ss}<0 & \Rightarrow
  &\frac{\sigma_{uu}}{1-\sigma_{uu}}>\frac{f_u}{1-f_u} \, ,
  &\frac{\sigma_{ll}}{1-\sigma_{ll}}< \frac{f_l}{1-f_l}
  &\textrm{and} & \frac{n_\textrm{ph}}{1+n_\textrm{ph}}<\frac{n_\textrm{b}}{1+n_\textrm{b}}
  &\Rightarrow & \frac{f_u}{1-f_u} <\frac{f_l}{1-f_l}\times \frac{n_\textrm{b}}{1+n_\textrm{b}}\\
 \end{array} \]
As the right hand-side needs to satisfy one of the three relations, we find equivalence for all relations. This provides the condition \eqref{EqSignQuantum}.

\end{widetext}

\section{Generalization for arbitrary systems with classical fields}\label{SecGeneralFermion}
The approach for the 2-level system used in the main article can be generalized to arbitrary fermionic systems with a time-periodic Hamiltonian
$\hat{H}_S(t)=\hat{H}_S(t+\tau)$ as characteristic for a classical optical field. This includes an arbitrary number of reservoirs, which are coupled to the system in the local form \eqref{EqLiouvillian}, where each reservoir $\alpha$ provides transitions to a unique level $\alpha$ of the system. Here, we assume that the system eventually reaches a steady state  $\hat{\rho}^\textrm{ss}(t)$ which is periodic with period $\tau$ following the external driving in $\hat{H}_S(t)$ after initial conditions died out due the dissipative terms in the quantum evolution. We define
\begin{eqnarray}
  \dot{U}^\textrm{av}_\alpha&=&\left\langle \textrm{Tr}\{\hat{H}_\textrm{S}(t)\mathcal{L}_\alpha[\hat{\rho}^\textrm{ss}(t)]\}
\right\rangle \\
\dot{N}^\textrm{av}_\alpha&=& \left\langle  \textrm{Tr}\{ \hat{a}_\alpha^\dag  \hat{a}_\alpha\mathcal{L}_\alpha[\hat{\rho}^\textrm{ss}(t)]\}\right\rangle\label{EqNav}
\end{eqnarray}
where the averaging 
  \[
  \langle f(t)\rangle=\frac{1}{\tau}\int_t^{t+\tau} f(t')\d t'
  \]
is taken over the common period of $\hat{\rho}^\textrm{ss}(t)$ 
and $\hat{H}_\textrm{S}(t)$. Thus $\dot{U}^\textrm{av}_\alpha$ and $\dot{N}^\textrm{av}_\alpha$ do not depend on time. This allows for the definition of effective energies
\begin{equation}
\tilde{E}_\alpha= \frac{\dot{U}^\textrm{av}_\alpha}{\dot{N^\textrm{av}_\alpha}}
\end{equation}
which are the average energies taken from the reservoir $\alpha$ per particle transferred into the system. It appears natural to apply  
these energies $\tilde{E}_\alpha$ for the energy dependence \eqref{EqfwithEtilde} of the occupation function $f_\alpha$ (and possibly also $\gamma_\alpha$, if the wide band limit is not applicable). This requires a self-consistent solution for the steady state of the dynamical equation \eqref{EqMasterEquation} with its input parameters. 

In the remaining part of this section, it is shown, that this procedure is thermodynamic consistent. The energy is conserved and the entropy production is semi-positive for the steady state averaged over a period. This follows concepts used in \cite{KosloffEntropy2013} for periodically driven systems in a related context.

The general expressions (\ref{EqEnergyLeadAlpha},\ref{EqWQquantumFull}) provide
\[
\frac{\d }{\d t} \textrm{Tr}\{\hat{H}_S(t)\hat{\rho}(t)\}=\sum_\alpha \dot{U}_\alpha+P_S\, .
\]
For the steady state the average change $\langle\frac{\d }{\d t} \textrm{Tr}\{\hat{H}_S(t)\hat{\rho}^\textrm{ss}(t)\}\rangle$ vanishes due to the integration over one period and we directly get $\sum_\alpha \dot{U}^\textrm{av}_\alpha+P_S^\textrm{av}=0$ showing that the average external power and energy currents from the reservoirs going into the system add up to zero. This is just energy conservation for the steady state.

The total entropy  $S_\textrm{tot}$ is given by the von Neumann entropy of the system and the reservoirs (changing by heat transfer  $\dot{U}_\alpha-\mu_\alpha\dot{N}_\alpha$ into the system), resulting in its temporal change
\[
\frac{\d S_\textrm{tot}}{\d t} =-k_B\frac{\d }{\d t} \textrm{Tr}\{\hat{\rho}(t)\ln\hat{\rho}(t)\}-
\sum_\alpha\frac{\dot{U}_\alpha-\mu_\alpha\dot{N}_\alpha}{T_\alpha}
\]
Upon averaging in the steady state we get
\[
\dot{S}^\textrm{av}_\textrm{tot}=\left\langle \frac{\d S_\textrm{tot}[\rho^\textrm{ss}(t)]}{\d t} \right\rangle =
-\sum_\alpha\frac{\tilde{E}_\alpha-\mu_\alpha}{T_\alpha}\dot{N}^\textrm{av}_\alpha
\]
Using Eq.~\eqref{EqNav} we find
\begin{equation}
\dot{S}^\textrm{av}_\textrm{tot}=k_B\left\langle\sum_\alpha \textrm{Tr}
\left\{\mathcal{L}_\alpha[\hat{\rho}^\textrm{ss}(t)]
\ln\hat{\rho}^\textrm{lt}\right\} \right\rangle \label{EqDotsSav}
\end{equation}
with the locally thermal operator
\[
\hat{\rho}^\textrm{lt}=\frac{1}{Z^\textrm{lt}}\prod_\alpha \exp\left(- \frac{\tilde{E}_\alpha-\mu_\alpha}{k_B T_\alpha}\hat{a}_\alpha^\dag  \hat{a}_\alpha \right)
\]
where the number $Z^\textrm{lt}$ renormalises the trace to unity (and drops out as $\textrm{Tr}\left\{\mathcal{L}_\alpha[\hat{\rho}]\right\}=0$).
As Eq.~\eqref{EqMasterEquation} gives
\[
\frac{\d }{\d t} \textrm{Tr}\{\hat{\rho}(t)\ln\hat{\rho}(t)\}=\sum_\alpha  \textrm{Tr}
\left\{\mathcal{L}_\alpha[\hat{\rho}(t)]\ln\hat{\rho}(t)\right\}\, ,
\]
we find for steady-state averaging
\[
0=k_B\left\langle \sum_\alpha  \textrm{Tr}
\left\{\mathcal{L}_\alpha[\hat{\rho}^\textrm{ss}(t)]\ln\hat{\rho}^\textrm{ss}(t)\right\}\right\rangle
\]
which we can subtract from  Eq.~\eqref{EqDotsSav} resulting in
\[
\dot{S}^\textrm{av}_\textrm{tot}=k_B\left\langle\sum_\alpha \textrm{Tr}
\left\{\mathcal{L}_\alpha[\hat{\rho}^\textrm{ss}(t)]\left(
\ln\hat{\rho}^\textrm{lt}-\ln\hat{\rho}^\textrm{ss}(t)\right)\right\} \right\rangle\, .
\]
As $\hat{\rho}^\textrm{lt}$ satisfies  $\mathcal{L}_\alpha[\hat{\rho}^\textrm{lt}]=0$ if the  occupations \eqref{EqfwithEtilde} are applied, we find $\dot{S}^\textrm{av}_\textrm{tot}\ge 0$ by Spohn's inequality \cite{SpohnJMathPhys1978,SpohnBook1978,LiPRE2017}. Thus, entropy production is positive semi-definite for the average steady state.

%

\end{document}